# Limited local electron-lattice coupling in manganites

D. Sánchez[1], M.J. Calderón[2], J. Sánchez-Benítez[3], A.J. Williams[3], J.P. Attfield[3], P.A. Midgley[1] and N.D. Mathur[1]*

[1]*Department of Materials Science, University of Cambridge, Pembroke Street, Cambridge CB2 3QZ, UK*
2 *Instituto de Ciencia de Materiales de Madrid (CSIC), Cantoblanco, 28049 Madrid, Spain.*
[3]*School of Chemistry and Centre for Science at Extreme Conditions, University of Edinburgh, West Mains Road, Edinburgh EH9 3JJ, UK*

*E-mail: ndm12@cam.ac.uk

(Pr,Ca)MnO$_3$ is the archetypal charge-ordered manganite, but in Pr$_{0.48}$Ca$_{0.52}$MnO$_3$ we find (using convergent-beam electron diffraction and dark-field images) that the superlattice period is locally incommensurate with respect to the parent lattice, and that the superlattice orientation possesses significant local variations. This suggests that local electron-lattice coupling never overwhelmingly dominates the rich physics of manganites, even in the most extreme scenarios that produce the largest colossal magnetoresistance effects.

In chemically single-phase perovskite manganites, there is widespread interest in the coexistence of magnetic and electronic phases over a wide range of length scales.[1,2] One of the predominant phases possesses a superlattice (below $T_s$) that is traditionally associated with charge order[3,4] (CO) arising due to strong local electron-lattice (Jahn-Teller) coupling.[5] The concept of strong coupling is entrenched in manganite physics because it is also considered[1] necessary for colossal magnetoresistance[6,7] (CMR), and the pronounced influence of strain on physical properties.[8] Here we investigate a member of the (Pr,Ca)MnO$_3$ family that epitomises[1,7,9-12] strong coupling. Surprisingly, we find that the local coupling is too weak to force the superlattice to be (i) locally commensurate or (ii) aligned with the parent lattice. This suggests that local coupling is of limited strength, thus extending the validity of a recent mean-field reinterpretation[13] cast purely in terms of magnetization and charge density.

In manganite compositions that show ferromagnetic metallic (FMM) manganites phases, the electron-lattice coupling is limited.[1] At low temperatures, the effective coupling is reduced by metallic screening, and even in the vicinity of the Curie temperature the high-field magnetoresistance effects are not colossal (e.g. just 4% at 5 T in[14] La$_{0.7}$Sr$_{0.3}$MnO$_3$, and 900% at 4 T in[15] La$_{0.75}$Ca$_{0.25}$MnO$_3$). The discovery[6] of CMR (10$^6$ % in 6 T) was made in a similar composition (La$_{0.67}$Ca$_{0.33}$MnO$_3$), but only in a sample where the microstructure had been altered by thin-film strain and annealing.

The largest CMR effects[7] (10$^{10}$ % at 5 T for Pr$_{0.67}$Ca$_{0.33}$MnO$_3$) in manganites arise at low temperatures in phases where it is believed[1] that strong electron-lattice coupling produces charge order. This phase is characterized by a superlattice (wavevector $\mathbf{q}\|\mathbf{a}^*$) seen in electron,[3,4] x-ray[16] and neutron[16,17] diffraction experiments. Originally this was interpreted[3,4] in terms of (200) planes containing the idealised cation Mn$^{4+}$, interspersed with orbitally ordered Mn$^{3+}$ planes (indexing the room temperature cell as orthorhombic *Pnma*). However, the key electron microscopy evidence for this description was overinterpreted and the nature of the modulations remains controversial.[18]

Recently we argued[18,19] against strong electron-lattice coupling in the modulated (so-called CO) phases of (La,Ca)MnO$_3$. Subsequently, our experimental findings were found to be reasonable in theoretical models where modulated phases are stabilized with small and even zero electron-lattice coupling.[13,20] However, one anticipates that our evidence[18,19] against strong coupling in (La,Ca)MnO$_3$ would not extend to a supposedly strong[1,7,9-12] CO manganite like (Pr,Ca)MnO$_3$. But surprisingly, in Pr$_{0.48}$Ca$_{0.52}$MnO$_3$ we find two independent pieces of evidence against strong local coupling. First, convergent-beam electron diffraction (CBED) data for Pr$_{0.48}$Ca$_{0.52}$MnO$_3$ shows that the superlattice is incommensurate with the parent lattice, cf. our La$_{0.48}$Ca$_{0.52}$MnO$_3$ data.[18] Second, we reveal that the superlattice orientation varies (i) locally in dark-field images as with (La,Ca)MnO$_3$,[21] and (ii) on a 100 nm length scale, where pronounced misorientations of **q** with respect to **a\*** are seen below 150 K, which we identify as the Néel temperature[9, 17] $T_N$ ($<T_s$).

Polycrystalline $Pr_{0.48}Ca_{0.52}MnO_3$ with micron-sized grains was prepared by repeated grinding, pressing and sintering of stoichiometric $Pr_6O_{11}$, $CaCO_3$ and $MnO_2$. Initially it was heated at 950°C for 12 hours to decarboxylate the $CaCO_3$, and then it was twice reground, repelleted and heated at 1350 °C for 4 days. X-ray powder diffraction in a Philips PW1050 diffractometer ($CuK_\alpha$) recorded only a single phase, and we infer that at least the grains of interest are chemically homogeneous given that we observe $q/a^*$ to be highly homogeneous. Electrically, the sample was found to be insulating and our resistance measurements suggest that 230 K $< T_s <$ 250 K as expected.[22] Electron transparency was achieved by conventional mechanical polishing and argon-ion thinning. A Philips CM30 transmission electron microscope was used to collect CBED patterns and 500 nm selected area diffraction (SAD) patterns. The microscope is equipped with a Gatan double-tilt liquid nitrogen ($LN_2$) stage with an effective base temperature of ~120 K. A Philips CM300 FEG TEM, equipped with a Gatan Image Filter and CCD camera, was used to collect energy-filtered 100 nm SAD patterns and dark-field images. This microscope supports Gatan sample stages with base temperatures of ~9.6 K (helium, data on warming only) and ~90 K (nitrogen).

Figure 1 repeats for $Pr_{0.48}Ca_{0.52}MnO_3$ the key CBED evidence[18] in $La_{0.48}Ca_{0.52}MnO_3$ which shows that the superlattice is locally incommensurate with a highly uniform period. This interpretation is forced by the failure of the CBED diffraction probe to record in any sampling the wavenumber $q/a^* = 0.5$ that would correspond to (orbitally ordered) alternating (200) planes[3] of $Mn^{3+}$ and $Mn^{4+}$ (we use these labels for convenience even though extreme charge separation is not expected[20]). This failure to observe these alternating planes is surprising in the traditional CO picture because at $x=0.52$ they should predominate, with the alternating pattern broken only by a few extra $Mn^{4+}$ planes for charge neutrality. These extra $Mn^{4+}$ planes would render the superlattice period ($2\pi/q$) non-uniform, and are expected[17, 18] on average every 9.6 nm (assuming[4, 23] a fine random mixture of two CO phase sub-unit species with $q/a^* = \frac{1}{2}$ and $\frac{2}{3}$) or 6.8 nm (assuming[18] that the sub-units are (002) planes of either $Mn^{3+}$ or $Mn^{4+}$). Our 3.6 nm CBED probe is therefore small enough to look between the putative extra $Mn^{4+}$ planes (but large enough to sample a few superlattice periods). In the CO picture, CBED should thus record $q/a^* = 0.5$ in most measurements. However, in ten separate grains where $q/a^*<0.5$ globally, our CBED probe never found $q/a^* = 0.5$. A typical CBED pattern appears in Fig. 1b. (Note that an eleventh grain had $q/a^*=0.5$ globally. This possibility was previously highlighted[24] but the present work shows that it is not a representative scenario.)

Figure 2 reveals that the orientation of the $Pr_{0.48}Ca_{0.52}MnO_3$ superlattice varies in space. We could not clearly resolve this in Fig. 1 because of averaging (Fig. 1a) and large-diameter reflections (Fig. 1b), but it is readily apparent (**q** not parallel to **a\***) in a typical 100 nm SAD pattern taken at 90 K (Fig. 2b). A dark-field image of the region from which this SAD pattern was taken shows variations in superlattice orientation on the nanoscale (Fig. 2c). At higher temperatures, there is no angular variation in 100 nm SAD patterns (180 K, Fig. 2d) but in fact the nanoscale variations persist (Fig. 2e). Angular variations on any length scale cannot be reconciled with dominant local coupling in general, or the traditional CO picture[3-5] in particular because it is based on (200) planes. Although angular variations could in principle arise via some complex arrangement of $Mn^{3+}$ and $Mn^{4+}$, they are most simply interpreted in a scenario where the electron-lattice coupling is limited due to competition with other degrees of freedom.

Figure 3 compares the magnitude $q/a^*$ (Fig. 3a) and orientation $\theta$ (Fig. 3b) of the superlattice wavevector (obtained from 100 nm SAD patterns taken from approximately the same location) with the bulk magnetization $M$ (Fig. 3c) measured using a Quantum Design SQUID. Below 150 K, (1) all but one value of $\theta$ is non-zero within error and the material is primarily antiferromagnetic,[9, 17] and (2) $q/a^* \approx 0.46$ varies little with temperature, but falls short of the expected[13] $(1-x)=0.48$ perhaps due to pinning[21] (which would also produce the local variations in superlattice orientation seen in Figs. 2c&e). Above 150 K, $q/a^*$ develops a strong temperature dependence (Fig. 3a), and all but three values of $\theta$ are zero within error (Fig. 3b). This transition in $\theta$ at 150 K likely represents the Néel transition in the sampled grain ($T_N$=170 K for $Pr_{0.5}Ca_{0.5}MnO_3$),[9, 17] and is consistent with the onset of the bulk antiferromagnetic transition on warming (Fig. 3c). Given that $\theta$ is only non-zero below $T_N$ we suggest that on a 100 nm measurement length scale, variations in superlattice orientation are intimately linked to the antiferromagnetic order (see later). We support this argument by noting that (1) $M$ (Fig. 3c) and $q/a^*$ (Fig. 3d) both show thermal hysteresis in approximately 100 K $< T <$ 200 K; (2) $q/a^*$ in 90 K CBED patterns is reduced by 2% when measured using a Lorentz lens and thus avoiding the 2 T objective field; and (3) below 40 K (Fig. 3c), the development of a small ferromagnetic component[10] (inset, Fig. 3c) is accompanied by weak evidence for a small reduction in $q/a^*$ (Fig. 3a).

It is surprising that in bulk $Pr_{0.48}Ca_{0.52}MnO_3$ we find the superstructure to be (1) locally incommensurate (Fig. 1b) and (2) mis-aligned with the parent lattice (Figs. 2b, 2c, 2e & 3b). (Pr,Ca)MnO$_3$ has hitherto been considered[1, 7, 9-12] to display the most extreme CO phases of manganites because $Pr^{3+}$ is smaller than $La^{3+}$,[25] because CMR effects are maximized,[7] and because there are no FMM phases at any doping.[10] However, our findings show that the superstructure cannot be interpreted so directly as a manifestation of extreme CO. If the local electron-lattice (Jahn-Teller) coupling does not dominate in as extreme a manganite composition as (Pr,Ca)MnO$_3$, its prominence must be diminished in manganite physics generally.

The magnetic correlations seen in Fig. 3 also require the local electron-lattice coupling to compete with other interactions rather than to dominate. The diffraction patterns in Figs. 2&3 suggest that $\theta$ is finite below $T_N$ and zero above, and Fig. 3b tends to confirm this. It is not known whether the local magnetic order in the antiferromagnetic state is locally commensurate as it might be in the modulated phases of (La,Ca)MnO$_3$,[26] but we assume commensurability which implies that the coupling is not as weak as we have previously suggested may be possible.[18, 19] Fig. 4 shows that the Jahn-Teller energy saving may be maximised if spatial variations in $\theta$ are associated with the presence of variable-width domains in a CE-type[27] antiferromagnetic configuration. In this cartoon, our typical 100 nm SAD value of $\theta \sim 0.5°$ suggests domains of width 22 nm (1° yields 10 nm). This width is less than our 100 nm measurement length scale, and consistent with the nanoscale variations in $\theta$ (Figs. 2c&e) that arise due to the Jahn-Teller coupling (and as seen in Fig. 2e persist above $T_N$). Note that large local variations in $\theta$ would be consistent in our interpretation (Fig. 4) with the presence of domain walls perpendicular to those depicted, as these produce a sign change in $\theta$.

Our demonstration of limited local coupling in manganites is not consistent with the extremely high electrical resistivities that might be expected in the CO picture. However, this expectation has no experimental validity. The modulated phases possess an activated resistivity with a gap of just $\sim 0.1$ eV, which corresponds to narrow-gap semiconducting behaviour.[28, 29] Note that the absolute resistivity of (Pr,Ca)MnO$_3$ ($10^4$ Ohm.cm at 100 K)[22, 30] is 5-6 orders of magnitude smaller than a good ferroelectric[31] which renders invalid the possibility[32] of ferroelectric measurements. In practice, measurements of resistance may yield significant overestimates of resistivity due to twins which are typically present even in single crystals.[33]

The 'electronically soft' reinterpretation[13] of manganites may be understood in the context of our surprise demonstration that the local electron-lattice coupling in manganites is not excessive. It remains a possibility that the local coupling is sufficiently weak[13, 18-20] for the modulated phases of manganites to be best described in a charge-density wave scenario,[18] rather than in the traditional CO interpretation. Indeed, some degree of electron itineracy is consistent with the reported resistivity data discussed above. Therefore the nature of the superlattice remains an open question.

We thank James Loudon for help with microscopy, and Peter Littlewood and Luis Brey for insight. This work was supported by an FP6 EU Marie Curie Fellowship (D.S.), the UK EPSRC, MAT2006-03741 and the Programa Ramón y Cajal (MEC, Spain).

**Figure captions**

FIG. 1. Electron diffraction patterns for a $Pr_{0.48}Ca_{0.52}MnO_3$ grain near 120 K. Solid lines indicate parent reflections separated by reciprocal lattice vector **a***. The superlattice (wavevector **q**) produces reflections at (a) $q/a^* = 0.4379(3)$ in a 500 nm selected-area diffraction (SAD) pattern, and (b) $q/a^* = 0.440(1)$ in a convergent beam electron diffraction (CBED) pattern (spot size ~ 3.6 nm). The dramatic reduction in the experimental length scale from (a) to (b) yields in effect no change in $q/a^*$ (the 0.5% difference falls within the small observed intragranular variation[18] in $q$). In particular, a lock-in to $q/a^* = 0.5$ (dashed lines) was not seen in all CBED patterns from this and other grains (with one exception that is irrelevant because the 500 nm SAD pattern found $q/a^* = 0.5$ for the entire grain). Therefore the superlattice is locally incommensurate with a highly uniform period.

FIG. 2. Local variations in the orientation of the superlattice in the same region of a $Pr_{0.48}Ca_{0.52}MnO_3$ grain. (a) schematic diffraction pattern showing the relation between parent lattice reflections (•), superlattice reflections (o) and the objective aperture (dashed line). Angles $\theta$ and $\phi$ are exaggerated for clarity. At 90 K we show (b) a 100 nm SAD pattern and (c) a dark-field image. At 180 K we show (d) a 100 nm SAD pattern and (e) a dark-field image. From (b) it is apparent that in a spatial average over 100 nm, the superlattice wavevector **q** contains a component along **c*** and is misaligned ($\theta \neq 0$) with respect to the parent lattice wavevector **a***. From (d) we see that this misorientation is absent at 180 K. Both dark-field images show interference fringes formed as a consequence of two nearby superlattice reflections (e.g. at **q** and **a***-**q**) falling within the objective aperture, and reveal that variations in superlattice misalignment arise on the nanoscale.

FIG. 3. Temperature dependence of electron diffraction and magnetization data for $Pr_{0.48}Ca_{0.52}MnO_3$. Values of (a) $q/a^*$ and (b) $\theta$ obtained from 100 nm SAD patterns taken from approximately the same region within a grain on warming. In (a) we present the 227 K and 9.6 K SAD patterns. As seen in (b) we find $\theta \neq 0$ (for all but one measurement) below ~150 K and (in all but three measurements) $\theta = 0$ above ~150 K. In (b) the magnitude of $\theta$ should only be interpreted to be zero or finite since the sampled area (i) contains nanoscale variations (Figs. 2c&e) and (ii) is subject to thermal drift. (c) the magnetization $M$ of a bulk sample on cooling and warming in 0.1 T reveals the expected[17] antiferromagnetic transition in 100-250 K. Thermal hysteresis is seen in both (c) $M$ and (d) $q/a^*$ obtained from 100 nm SAD patterns taken within a grain on warming and cooling (in a nitrogen stage). The expected[10] ferromagnetic component that develops below 40 K shows magnetic hysteresis (inset of (c)), and there is weak evidence for a small concomitant decrease of $q/a^*$. Error bars in $q/a^*$ and $\theta$ indicate the standard deviation of the values extracted from the diffraction pattern.

FIG. 4. Variations in superlattice misalignment from competition between locally uniform charge modulation and limited local electron-lattice (Jahn-Teller) coupling. Dots represent Mn atoms in this (010) cross-section. We assume CE-type antiferromagnetism, where ferromagnetic zig-zag chains of Mn (shaded) are antiferromagnetically coupled to neighbouring chains, and the valence electron density is larger on the bridge sites (dark) than the corner sites (light). The locally incommensurate superlattice is represented by parallel lines that denote maxima in charge density, and the on-site Jahn-Teller energy saving is maximised when these lines fall as close as possible to bridge sites. Within an antiferromagnetic domain, the charge-density maxima fall out of sync with the bridge sites when the superlattice (wavenumber $q$) is (a) collinear or (b) misaligned with respect to the orthorhombic axis **a**. (c) the presence of suitably spaced antiferromagnetic domain walls and variations in $\theta$ permit the charge-density maxima of a non-collinear modulation to remain near the bridge sites everywhere.

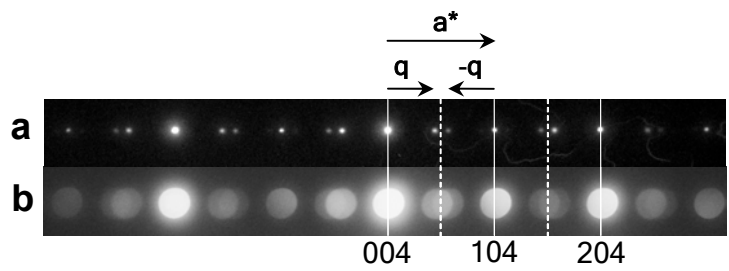

Fig. 1

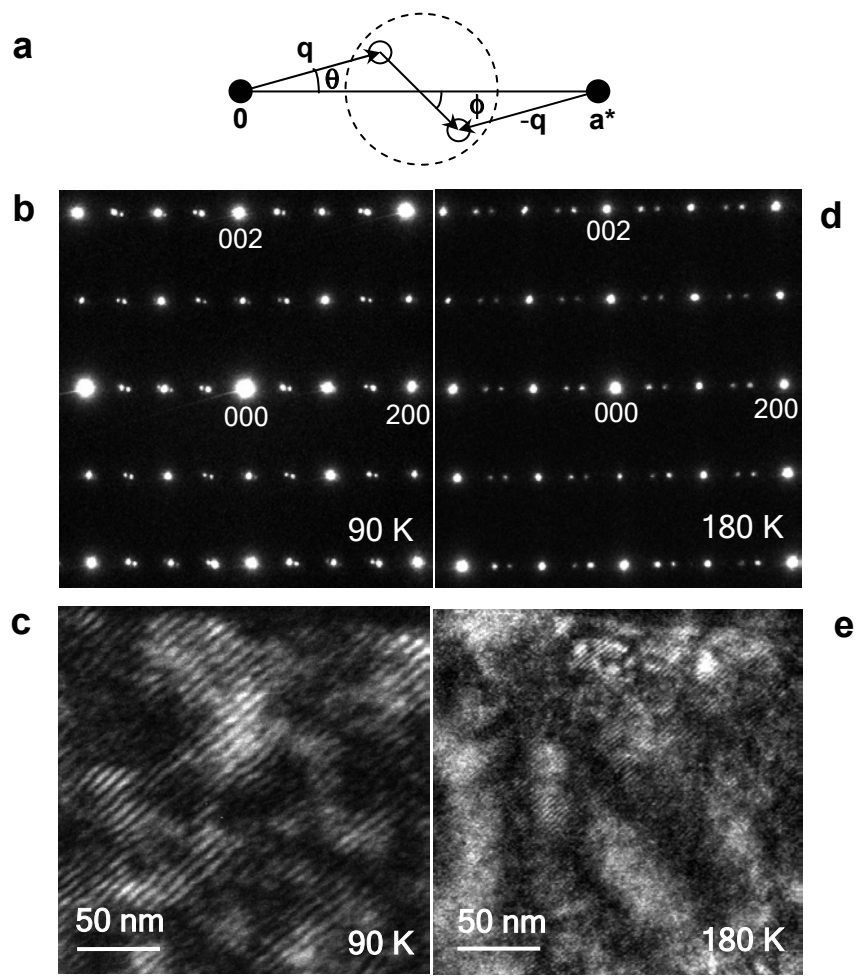

Fig. 2

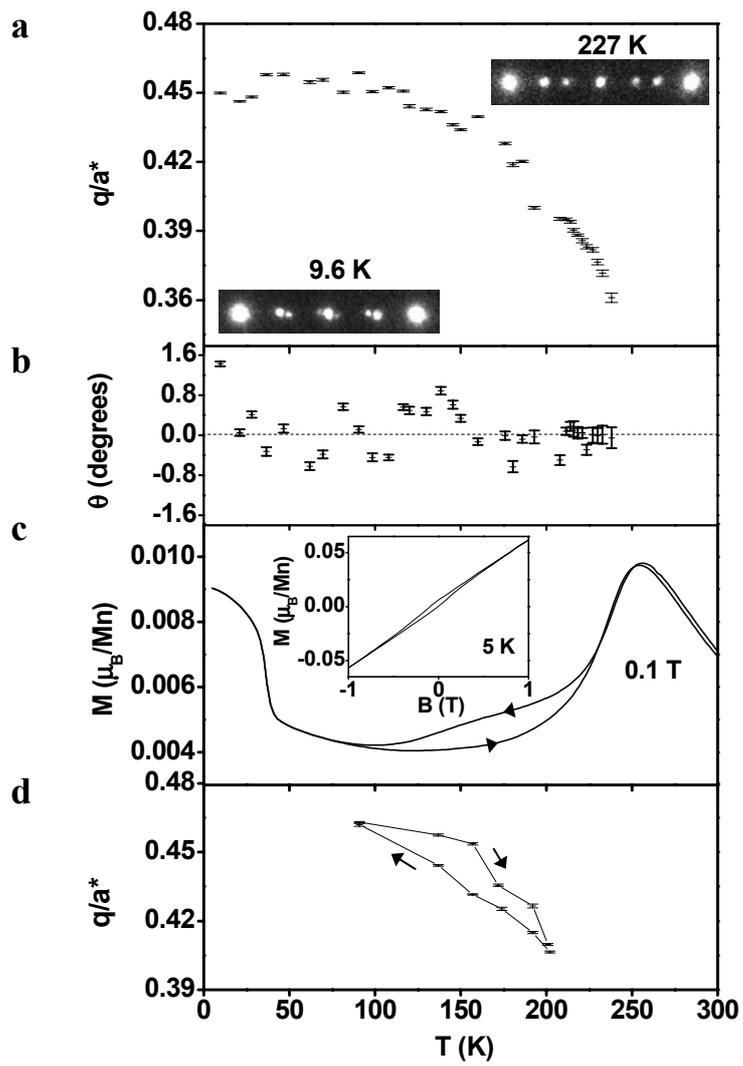

Fig. 3

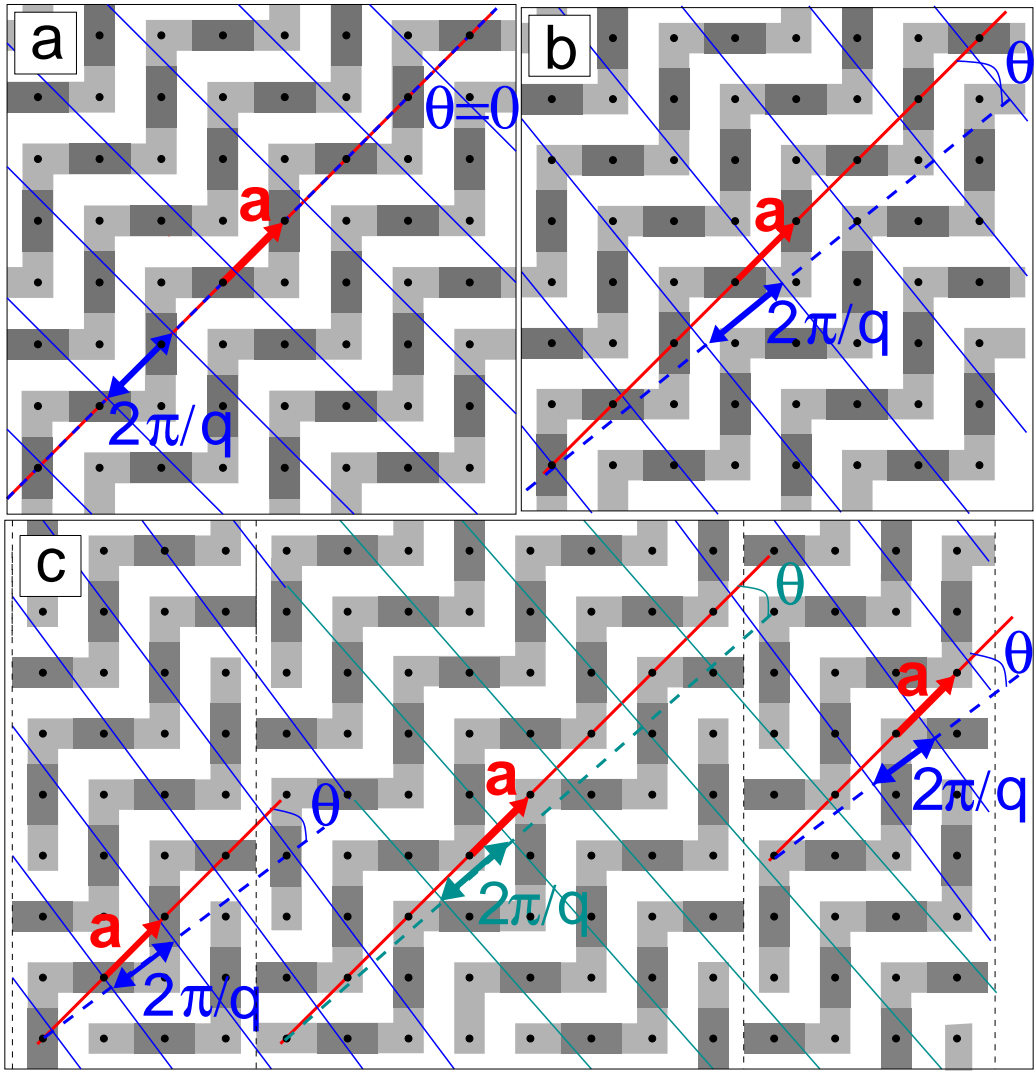

Fig. 4